\documentclass[aps,showpacs,superscriptadress,preprint]{revtex4}
\usepackage{graphicx}
\usepackage{amsmath}

\begin{document}
\title{Quark deconfinement phase transition in nuclear matter for improved quark mass
       density-dependent model}

\author{
Chen Wu$^{1}$\footnote{wuchenoffd@gmail.com}, Wei-Liang Qian$^{2}$,
Yu-Gang Ma$^{1}$, Guo-Qiang Zhang$^{1}$, and Sanjeev Kumar$^{1}$ }
 \affiliation{
\small 1. Shanghai Institute of Applied Physics, Chinese Academic of Sciences, Shanghai 201800, China \\
\small 2. Universidade de Ouro Preto, Ouro Preto, 35400-000,
Brazil\\ }

\begin{abstract}
The improved quark mass density-dependent (IQMDD) model, which has
been successfully used to describe the properties of both infinite
nuclear matter and finite nuclei, is  applied to  investigate the
properties of quark deconfinement phase transition. By using the
finite-temperature quantum field theory, we calculate the finite
temperature effective potential and extend the IQMDD model to finite
temperature and finite nuclear matter density. The critical
temperature and the critical  density of nuclear matter are given
and the QCD phase diagram is addressed. It is shown that this model
can not only describe the saturation properties of nuclear matter,
but also explain the quark deconfinement phase transition
successfully.
\end{abstract}

\pacs{11.10.Wx, 12.39.Ba, 25.75.Nq} \maketitle

 Owing to
the nonperturbative nature of quantum chromodynamics (QCD) in low
energy regions, it is very difficult to study nuclear system by
using QCD directly. Phenomenological models reflecting the
characteristic of the strong interaction are widely used in the
studying of the properties of hadrons and nuclear matter. The
quark-meson coupling (QMC) model suggested by Guichon \cite{QMC} is
a famous hybrid quark meson model, which can describe the saturation
properties of nuclear matter and many other properties of finite
nuclei successfully [2,3]. In this model, the nuclear system was
suggested as a collection of MIT bag, vector $\omega$-meson and
scalar $\sigma$-meson, and the interactions between quarks and
mesons are limited within the MIT bag regions due to the quark
cannot escape from the MIT bag.

Although the QMC model is successful for describing the physical
properties of a nuclear system, two shortcomings arise when one uses
this model to discuss some physical topics. First, the MIT  bag
model  is a permanent quark confinement model and  its boundary
condition cannot be destroyed by temperature and nuclear matter
density. The Second difficulty arises from the MIT boundary
condition. If we hope to do the nuclear many-body calculations
beyond mean field approximation (MFA) by quantum field theory, it is
essential to find the free propagators of quark, $\sigma$ meson and
$\omega$ meson, respectively. However, the constraint of the MIT bag
boundary condition presents obstacles to get the corresponding
propagators in free space. Because the interactions between quarks
and mesons are limited within the bag regions, and multireflection
of quarks and mesons by the boundary must  be taken into account
\cite{multi-reflection}.

In order to keep the quark confined property and  give up the MIT
boundary condition,  a new quark-meson coupling model based on a
quark mass density-dependent (QMDD) model was presented in our
previous papers [5,6]. According to the QMDD model,  the masses of
u, d and s quarks (and the corresponding antiquarks) satisfy:
\begin{eqnarray}
m_{q} = \frac{B}{3n_{B}}(q = u,d,\bar u,\bar d),   \label{mq1}
\end{eqnarray}
\begin{eqnarray}
m_{s,\bar s } = m_{s0}+\frac{B}{3n_{B}},     \label{mq2}
\end{eqnarray}
where $n_{B}=\frac{1}{3}(n_u+n_d+n_s)$ is the baryon number density,
with $n_u, n_d, n_s$ representing the density of the $u$ quark, $d$
quark, and $s$ quark,
 $m_{s0}$ is the current mass of the
strange quark, and  $B$ is the vacuum energy density inside the bag.
As was explained in Refs. \cite{QMDD}, the ansatz Eqs. (1) and (2)
corresponds to a quark confinement hypothesis because when
$V\rightarrow \infty, n_B \rightarrow 0$ and $m_q \rightarrow
\infty$, it prevents the quark from going to infinity or to very
large regions. The large volume means that the baryon density is
small. This mechanism of confinement can be mimicked through the
requirement that the mass of an isolated quark become infinitely
large so that the vacuum is unable to support it \cite{QMDD-2}. This
is just the physical picture given by Eqs. (1) and (2). In fact,
this confinement mechanism is very similar to that of the MIT bag
model \cite{QMDD-3}. But the advantage of the QMDD model is that it
does not need to introduce a quark confined boundary condition like
that of the MIT bag model.

But  the QMDD model still has two shortcomings: (1) it is still an
ideal quark gas model. No interactions between quarks exist except
for a confinement ansatz, equation (1) and (2). (2) it  cannot
describe the quark deconfinement phase transition and predict a
correct phase diagram as that given by lattice QCD. The reason is
that the temperature $T$ tends to infinite when density $n_B
\rightarrow$  0. This result can easily be understood if we notice
the basic hypothesis Eqs. (1) and (2) of the QMDD model, the quark
masses are divergent when $n_B \rightarrow$ 0. To excite an infinite
weight particle, one must prepare to pay the price for infinite
energy, i.e., infinite temperature. It means that the confinement in
the QMDD model is still permanent. To overcome the difficulty (2),
we have introduced a new ansatz that the vacuum density $B$ is a
function of temperature $T$ \cite{Zhang Yun}:
\begin{eqnarray}
B(T) &=& B_0 [1- (T/T_C)^2], 0\leq T \leq T_C, \\
B(T) &=& 0,  T > T_C.
\end{eqnarray}
and extended the QMDD model to a quark mass density- and
temperature- dependent (QMDTD) model  \cite{Zhang Yun}. This QMDTD
model has been employed to discuss the properties of strange quark
matter \cite{Zhang Yun-1}, the dibaryon system \cite{Zhang Yun-2}
successfully.

To improve the shortcoming (1) of the QMDD model and to avoid the ad
hoc ansatz of QMDTD model,  we introduced the $\omega$-meson and
$\sigma$-meson in the QMDD model to mimic the repulsive and
attractive interactions between quarks in Refs. [5,6]. Following the
treatment of Friedberg and Lee [13,14], the interaction between
quarks and the nonlinear $\sigma$ field forms a Friedberg-Lee bag
\cite{Wu-PRC-1}. In Ref. \cite{Wu-PRC-1}, we found the wave
functions of the quark ground state and the lowest one-quark excited
states. By using these wave functions, we calculated many physical
quantities such as root-mean-square radius, the magnetic moment of
nucleon to compare with experiments and come to a conclusion that
this IQMDD model is successful to explain the properties of nucleon.
Then under mean field approximation, we employ the IQMDD to
investigate the physical properties of nuclear matter.
 Instead of the MIT bag, after introducing the nonlinear
interaction of $\sigma$-mesons and qq$\sigma$ coupling, we construct
a Friedberg- Lee soliton bag in nuclear system.  The quark and
$\sigma$- meson coupling and the quark and $\omega-$ meson coupling
are introduced to mimic the attractive and the repulsive
interactions between quarks in this model.
 It has been shown that our model can successfully describe the saturation properties, the
equation of state, the compressibility and the effective nucleon
mass of symmetric nuclear matter \cite{Wu-PRC-2} and finite nuclei
\cite{Wu-JPG-3}.

We hope to emphasize that there are two basic differences between
the IQMDD model and the usual QMC model. First, unlike in the QMC
model, we do not need an MIT bag for the nucleon in the IQMDD model.
The constraint of the MIT bag boundary condition disappears in our
formulas because the quark confinement mechanism has been
established in Eqs. (1) and (2). Second, the interaction between the
quark and the scalar meson is limited in the bag regions for the QMC
model, while for the IQMDD model this interaction is extended to the
whole free space. So it is convenient to write down the free
propagators of quarks and mesons.

In Ref. [16,17], we extended this model to finite temperature and
studied its soliton solution by means of the finite temperature
quantum field theory. The critical temperature of quark
deconfinement $T_C$ and the function of temperature-dependent bag
constant $B(T)$ are found as an output. But the the deconfinement
properties of the IQMDD model at finite density of nuclear matter
and the QCD phase diagram have not yet been studied. We would like
to  study these problems in this paper. We will show that we can get
reasonable quark deconfinement critical temperature and critical
density of nuclear matter. Moreover the QCD phase diagram can also
be given in the IQMDD model phenomenologically.

Before giving the main  formulas of the IQMDD model, we should
emphasize that the quark deconfinement phase transition and QCD
phase diagram have  attracted a lot of interest for many years. many
theoretical  efforts have been recently and less recently carried
out, on the study of the hadronic phase structure and the
deconfinement phase transition by using the Nambu-Jona-Lasinio model
and some its extensions in the framework of finite temperature field
theory [19-30]. Among them, some perform a very detailed
investigation of the QCD phase structure and other ones give precise
numerical estimates for the deconfinement temperature and size of
hadrons. These publications is very interesting and also give us
lots of important information for the QCD phase structure.

  The effective Lagrangian density of the IQMDD model is given by
\begin{eqnarray}
\mathcal{L}=\bar\psi[{i}\gamma^\mu\partial_\mu-
m_{q}- g_\sigma^q \sigma-g_\omega^q \gamma^{\mu}\omega_{\mu}]\psi \hskip 0.8in \nonumber\\
+\frac{1}{2}\partial_{\mu}\sigma \partial^{\mu}\sigma -U(\sigma)
-\frac{1}{4}F_{\mu\nu}F^{\mu\nu} + \frac{1}{2}m_{\omega}^2
\omega^\mu \omega_\mu,
\end{eqnarray}
where
$F_{\mu\nu}=\partial_{\mu}\omega_{\nu}-\partial_{\nu}\omega_{\mu}$,
$\psi$ represents the quark field, $m_{q}=\frac{B}{3n_{B}}$ is the
mass of $u(d)$ quark, the $\sigma$  and $\omega$ fields are not
dependent on time, $g_\sigma^q$ is the coupling constant between the
quark field $\psi$ and the scalar meson field $\sigma$, $g_\omega^q$
is the coupling constant between the quark field $\psi$ and the
vector meson field $\omega_{\mu}$, $U(\sigma)$ is the self
interaction potential for $\sigma$ field. The potential field
$U(\sigma)$ is chosen as \cite{T.D. Lee}
\begin{eqnarray}
U(\sigma)=\frac{a}{2!}\sigma^2+\frac{b}{3!}\sigma^3+\frac{c}{4!}\sigma^4+B,
\end{eqnarray}
\begin{equation} \label{e7}
b^2 > 3ac.
\end{equation}
The condition (\ref{e7}) ensures that the absolute minimum of
$U(\sigma)$ is at $\sigma = \sigma_{v} \ne 0$. The potential
$U(\sigma)$ has two minima: one is the absolute minimum $\sigma_v$,
\begin{eqnarray}
\sigma_v=\frac{3|b|}{2c}\left[1+\left[1-\frac{8ac}{3b^2}\right]^{\frac
1 2}\right],
\end{eqnarray}
it  corresponds to the physical  vacuum,  and the other is at
$\sigma_0=0$, it represents a metastable local false vacuum. We take
$U(\sigma_v)=0$ and the bag constant $B$ can be expressed as
\begin{eqnarray}
-B=\frac{a}{2!}\sigma^2_v+\frac{b}{3!}\sigma^3_v+\frac{c}{4!}\sigma^4_v.
\end{eqnarray}

In order  to study the deconfinement phase transition, we turn to
extend IQMDD model to finite temperature.   The appropriate
framework is finite temperature quantum field theory. The finite
temperature effective potential plays a central role within this
framework. Under the mean field approximation, the $\omega$ meson
field operator can be replaced by its expectation value in symmetric
nuclear matter \cite{RMF}, \begin{eqnarray}
 \bar\omega_\mu = \delta_{\mu 0}
\bar\omega = \delta_{\mu 0} \frac{3g_\omega^q}{m_\omega^2} \rho_B.
\end{eqnarray}

Using the method of Dolan and Jackiw \cite{Dolan}, up to one-loop
approximation, the effective potential reads:
\begin{eqnarray}\label{potential0}
V(\sigma;\beta;  V_\omega)=U(\sigma)+
V_B(\sigma;\beta)+V_F(\sigma;\beta;  V_\omega),
\end{eqnarray}
where
\begin{eqnarray}
V_\omega= g_\omega^q \bar\omega= 3
\frac{{g_\omega^q}^2}{m_\omega^2}\rho_{B}
\end{eqnarray}
is the contribution of $\omega$-field, $\beta$ is the inverse of the
temperature, $\rho_{B}$ is the density of nuclear  matter and
\begin{eqnarray}
V_B(\sigma;\beta)=\frac{1}{2\pi^2 \beta^4} \int^{\infty}_0 dx x^2
\mathrm{ln} \left( 1-e^{-\sqrt{(x^2+\beta^2 m_{\sigma}^2)}} \right),
\end{eqnarray}
\begin{eqnarray}
V_F(\sigma;\beta;  V_\omega)=-12\sum_n \frac{1}{2\pi^2 \beta^4}
\int^{\infty}_0 dx x^2 \mathrm{ln} \left( 1+e^{-(\sqrt{(x^2+\beta^2
m_{q}^2)}-\beta V_{\omega})} \right),
\end{eqnarray}
where the minus sign of Eq. (14) is the consequence of Fermi-Dirac
statistics. The degenerate factor 12 comes from: 2(particle and
antiparticle), 2(spin), 3(color). $m_{\sigma}$ and $m_q$ are the
effective masses of the scalar field $\sigma$ and the quark field
respectively which can be found in Ref. [16,17].

We see from Eqs. (11-14) that the scalarlike interaction $g_\sigma^q
\psi^+\sigma\psi$ gives contribution for effective masses of quark
and $\sigma$\- meson and then forms a confined soliton bag, while
the vectorlike interaction $g_\omega^q
\psi^+\gamma^{\mu}\omega_{\mu}\psi$ gives contribution to an
effective chemical potential of quarks. In fact, this finite
temperature effective potential for FL model had been calculated by
many others authors [16,17].

In the FL bag  model, the finite-temperature vacuum energy density
$B(T)$ is defined as
\begin{eqnarray}\label{bag}
B(\beta;  V_\omega)=V(\sigma_0;\beta; V_\omega)-V(\sigma_v;\beta;
V_\omega).
\end{eqnarray}
It is the difference   between the values in  the perturbative false
vacuum state and the values in the physical real vacuum state of the
finite temperature effective potential. At critical density of
nuclear matter and  critical temperature  of quark deconfinement
phase transition, $B$ equal to zero: $B(T, \rho_B)=0$.

Before numerical calculation, let us discuss the parameters in IQMDD
model. First, we choose $m_\omega=783$ MeV, $m_\rho=770$ MeV and
$m_\sigma=509$ MeV as that of Ref. \cite{Wu-PRC-2}. Fixing the
nucleon mass $M_N=939$ MeV, we get $B=174$ MeV fm$^{-3}$. Obviously,
the behaviors at the saturation point must be explained  for a
successful model. It reveals that nuclear matter saturates at a
density $\rho_0=0.15$ fm$^{-3}$  with a binding energy per particle
$E/A= -15$ MeV at zero temperature, and the compression constant to
be about $K(\rho_0)=210$ MeV. Therefore we fixed $ g_\omega^q=2.44,
g^q_\sigma=4.67, b=-1460$ MeV to explain above data \cite{Wu-PRC-2}.

Now we turn to investigate the effective potential $V(\sigma; \beta;
V_\omega)$. The effective potential at finite temperature and finite
nuclear matter density can be obtained by numerical calculations
using the set of Eqs. (11-14). Then it is convenient to investigate
the temperature and the nuclear matter density dependence of the bag
constant. The bag constant in the IQMDD model is defined as Eqs.
(15). The upper panel of Fig. 1 shows  the temperature dependent
behavior of the bag constant at various nuclear matter densities.
One can see clearly that the bag constant decreases continuously
with increasing temperature at fixed nuclear matter density. At
critical temperature $T_C$, the bag constant equals to zero and the
deconfinement phase transition begins to occur. The lower panel of
Fig. 1 shows  the nuclear matter density dependence of the bag
constant at different temperatures. Similar to the upper panel of
Fig. 1, one can see the deconfinement phase transition begin to take
place at a critical
 density  for a given value of  temperature.

The critical behavior of nuclear matter in the $T$ - $\rho_B$ phase
diagram calculated using the IQMDD model is described  in Fig. 2.
The critical line corresponds to the phase transition  between
confined and deconfined phases. In this nuclear model, our
calculation shows that the critical density of nuclear matter for
quark deconfinement is about  ten times the saturation density of
the nuclear matter.   The phase diagram shown in Fig. 2, based on
the phenomenological Friedberg-Lee model, predicts a first-order
deconfinement phase transition for the full phase diagram. This is
due to at the critical $T$ and $\rho_B$ the two vacuum states,
namely the perturbative vacuum state and the physical vacuum state,
appear in the same energy. This result differs from the predictions
based on lattice gauge theory where more complicated phase diagram
on the $T$ and $\rho_B$ plane has been obtained, giving more
fruitful behaviors of the QCD phase [20,21]. For example, at finite
temperature and zero chemical potential, the phase transition could
be of a second order; moreover, it is predicted that there may exist
a critical point along the critical line.

In the end of paper, we have provided a comparison of the results
obtained from the IQMDD model with the recent results obtained in
the framework of effective four fermion interaction theories. The
recent efforts [21-30] have made  very fruitful progress on the QCD
phase structure  including finite-size hadron and magnetic effects.
Their theoretical frame of field theory is the imaginary-time
temperature field theory, same as the IQMDD model. As the IQMDD
model has lack of chirality, we can not predict critical point along
the critical line. However, we should emphasize that the IQMDD model
can not only explain the quark deconfinement phase transition, but
also describe the saturation properties of the nuclear matter
successfully. In the IQMDD model, the symmetry nuclear matter is
treated as uniform and unlimited,  the vectorlike interaction
$g_\omega^q \psi^+\gamma^{\mu}\omega_{\mu}\psi$ between
 quarks and $\omega$ meson gives contribution to an effective chemical potential of
quarks, and the expected value of $\omega$ meson is connected with
the baryon density. Our numerical results shows that the deconfining
baryon density at zero temperature is about 1.63 fm$^{-3}$ which is
very close to the  value predicted by the lattice QCD theory.
Meanwhile, the deconfining temperature at zero baryon density is 139
MeV which is also reasonable compared to 170 MeV given by lattice
QCD theory \cite{Lattce QCD}. Recently, the Science journal
\cite{science} report that the deconfining  critical temperature for
the QCD phase transition at zero baryon density is about 175 MeV.

In conclusion, we investigate the quark deconfinement phase
transition at finite temperature and finite density of  symmetric
nuclear matter in the IQMDD model. We would like to emphasize that
this is the basic  advantage of IQMDD model, because the saturation
properties can not only be explained
 by IQMDD model but also by QMC model. The reason for quark
 deconfinement can be explained by IQMDD model is that
MIT boundary constraint has been dropped out   and interactions
between quark and mesons have been extended to the whole space.
Instead of the MIT bag in QMC model, a FL soliton bag is introduced
in the IQMDD model, which makes it possible to discuss the quark
deconfinement phase transition. The spontaneously breaking symmetry
of nonlinear $\sigma$ field is restored and the soliton bag will
disappear at critical condition.
\newline

The author  wish to thank Professor  Ru-keng Su for useful
correspondences. This work is supported by the National Natural
Science Foundation of China (Grants 11105072 and 11035009).

\begin{figure}[tbp]
\includegraphics[width=14cm,height=20cm]{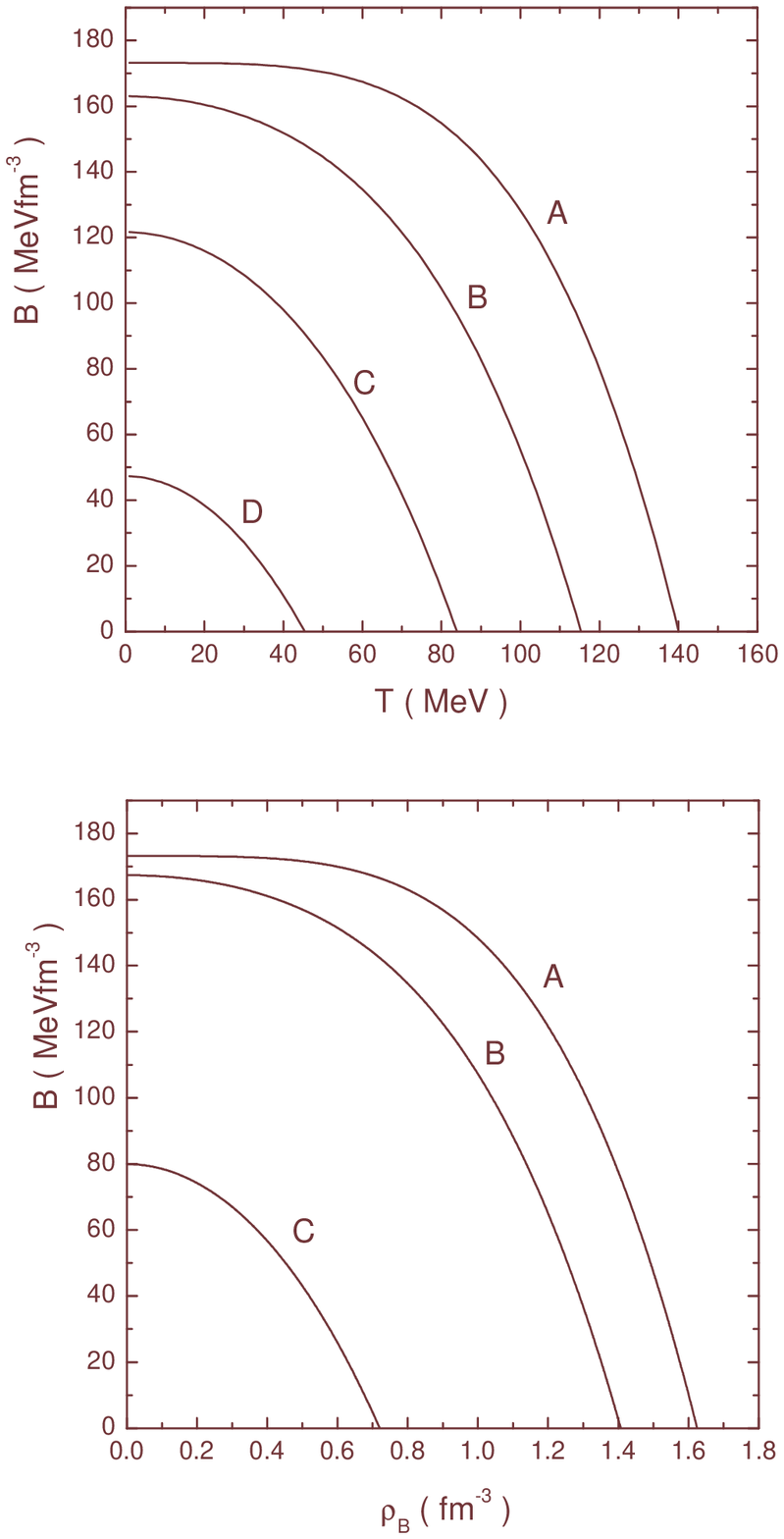}
\caption{(upper panel) Bag constant vs temperature at different
nuclear matter density. A: $\rho_B = 0.0$ fm$^{-3}$, B: $\rho_B =
0.8 $ fm$^{-3}$, C: $\rho_B = 1.2$ fm$^{-3}$, D: $\rho_B = 1.5$
fm$^{-3}$; (lower panel) Bag constant vs nuclear matter density at
different temperature. A: $T = 0$ MeV, B: $T = 60$ MeV, C: $T = 120$
MeV.}
\end{figure}

\begin{figure}[tbp]
\includegraphics[width=14cm,height=20cm]{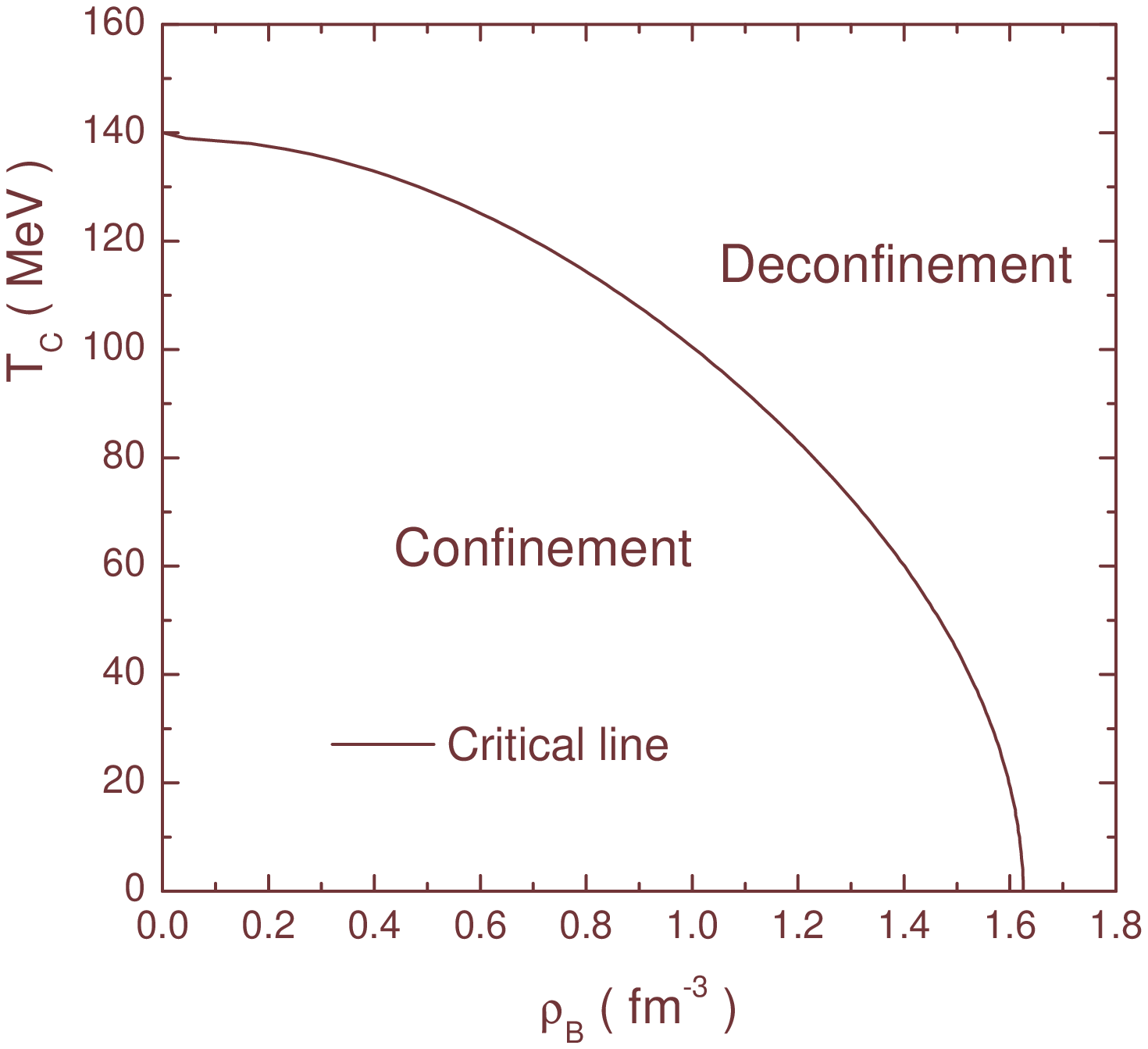}
\caption{Phase diagram of deconfinement in the IQMDD  model. }
\end{figure}

\end{document}